\documentclass[reprint,tightenlines,superscriptaddress,prd,nofootinbib,eqsecnum,showpacs]{revtex4}

\usepackage{amssymb}
\usepackage[dvips]{graphicx}
\def\dd{{\rm d}}

\def\_#1{^{}_{#1}}

\def\beq{\begin{equation}}
\def\eeq{\end{equation}}
\def\bea{\begin{eqnarray}}
\def\eea{\end{eqnarray}}
\usepackage{enumerate}

\begin{document}

\title{Anisotropic evolution of 5D Friedmann-Robertson-Walker spacetime}

\author{Chad A. Middleton}
\email{chmiddle@coloradomesa.edu}
\affiliation{Department of Physical and Environmental Sciences, Colorado Mesa University  (formerly Mesa State College), Grand Junction, Colorado 81501, USA}

\author{Ethan Stanley}\email{stanlee@purdue.edu}
\affiliation{Department of Physics, Purdue University, West Lafayette, Indiana 47907, USA}

\date{\today}

\begin{abstract}
We examine the time evolution of the five-dimensional Einstein field equations subjected to a flat, anisotropic Robertson-Walker metric, where the 3D and higher-dimensional scale factors are allowed to dynamically evolve at different rates.   By adopting equations of state relating the 3D and higher-dimensional pressures to the density, we obtain an exact expression relating the higher-dimensional scale factor to a function of the 3D scale factor.  This relation allows us to write the Friedmann-Robertson-Walker field equations exclusively in terms of the 3D scale factor, thus yielding a set of 4D effective Friedmann-Robertson-Walker field equations.    We examine the effective field equations in the general case and obtain an exact expression relating a function of the 3D scale factor to the time.   This expression involves a hypergeometric function and cannot, in general, be inverted to yield an analytical expression for the 3D scale factor as a function of time.  When the hypergeometric function is expanded for small and large arguments, we obtain a generalized treatment of the dynamical compactification scenario of Mohammedi [Phys.Rev.D {\bf65}, 104018 (2002)] and the 5D vacuum solution of Chodos and Detweiler [Phys.Rev.D {\bf21}, 2167 (1980)], respectively.   By expanding the hypergeometric function near a branch point, we obtain the perturbative solution for the 3D scale factor in the small time regime.  This solution exhibits accelerated expansion, which, remarkably, is independent of the value of the 4D equation of state parameter $w$.  This early-time epoch of accelerated expansion arises naturally out of the anisotropic evolution of 5D spacetime when the pressure in the extra dimension is negative and offers a possible alternative to scalar field inflationary theory.   


\end{abstract}\maketitle

\section{Introduction}

In the early 20th century, Theodor Kaluza and Oskar Klein attempted to unify the forces of electromagnetism and gravitation into a more fundamental description of nature by introducing extra dimensions into general relativity \cite{KK}.  Since the onset of Kaluza-Klein theory, the notion of incorporating extra dimensions into general relativity has remained as a potentially viable alternative to traditional four dimensional (4D) theories in accounting for various physical phenomena.
This addition of extra dimensions into physical theory has been more actively pursued in the last several decades (for a review, see Applequist, Chodos, and Frend \cite{Apple}).  The advent of string theory as a possible candidate to offer a consistent quantum description of gravity has been primarily responsible for fueling this activity in the research of extra dimensional scenarios as superstring theory has the strict requirement of residing in a 10D spacetime. Traditionally, the six extra spatial dimensions predicted by string theory have been treated as compactified objects, as are the dimensions of Kaluza-Klein theory. The discovery of D(irichlet)-branes as fundamental extended objects has provided a new possible explanation for the hidden extra spatial dimensions and has given rise to several different braneworld scenarios where the extra dimensions are large \cite{Large}. 

Recent observations of type Ia supernovae \cite{Accel} and measurements of the anisotropies of the cosmic microwave background \cite{CMB} indicate that our universe is, on the largest of scales, spatially flat, homogeneous, isotropic, and currently undergoing accelerated expansion.  This late-time epoch of accelerated expansion can be explained within the framework of the 4D standard model of cosmology by hypothesizing a dark energy component to the total energy density of the universe.  This dark energy is characterized by a negative pressure equation of state $p=w\rho$, where the equation of state (EoS) parameter $w$ lies near $-1$.  Dark energy can possibly be understood as arising from quantum fluctuations of the vacuum.  If this vacuum energy does in fact exist, it would dominate over radiation and matter in the late universe and seems to be a perfect candidate for dark energy.  Quantum theory, however, predicts a value far different from that which is observed \cite{CosCon}.  This inconsistency between theory and observation is known as the cosmological constant problem and remains as one of the greatest problems in theoretical physics today. 
  
A universe that is characterized to be spatially flat, homogeneous, and isotropic 
can be neatly explained through the phenomenon of early-time inflation \cite{Inflation, Liddle}.  Single field inflationary theory is characterized by an early-time accelerated expansion that is driven by a spatially uniform scalar field.  This scalar field is characterized by having a negative pressure when the scalar field ``slowly rolls" down a potential energy curve.  This slow roll allows the potential energy term to dominate over the kinetic energy term and gives rise to the required negative pressure, if one is to demand an epoch of accelerated expansion.  The end of this early-time inflationary epoch is reached when the slow-roll conditions are violated and is followed by an epoch of reheating.   Scalar field inflationary theory is incomplete, however, as the potential energy function is left undetermined by the theory.  In addition to this shortcoming, observational evidence of the existence of a fundamental scalar particle is nonexistent, although the search is on for the elusive Higgs boson!

Mohammedi offered an alternative explanation to dark energy being responsible for driving the late-time accelerated expansion of the universe by incorporating extra dimensions into Friedmann-Robertson-Walker (FRW) cosmology \cite{Mohammedi}.  There he considered the Einstein field equations in $D=d+4$ spacetime dimensions subjected to a flat, anisotropic Robertson-Walker metric, where the 3D scale factor is allowed to evolve at a rate different from that of the scale factor of the higher-dimensional space.  By assuming that the higher-dimensional scale factor dynamically evolves at a rate inversely proportional to a power of the 3D scale factor, Mohammedi showed that the $D$-dimensional FRW equations reduce precisely to a 4D form, once an \textit{effective pressure} is defined.  This effective pressure is a linear combination of the density and the 3D and higher-dimensional pressures.  It was noted in \cite{Mohammedi} that this effective pressure can be negative for positive values of the density and the pressures and can drive the late-time accelerated expansion of the universe, without the need for dark energy.  The work of \cite{Mohammedi} has been extended and studied in \cite{others, Pan}.  
The aforementioned findings of \cite{Mohammedi} were reached on the assumption of dynamical compactification of an inverse power law form.  It would be intriguing if this specific form of dynamical compactification arises naturally from a general treatment of the higher-dimensional FRW equations.

Here, we consider the five-dimensional Einstein field equations subjected to an anisotropic Robertson-Walker metric, where the scale factor for the 3D spatial manifold is allowed to evolve at a rate different, in general, from that of the scale factor of the extra dimension.  We additionally assume that the matter-energy content of the 5D universe is that of a perfect fluid, where the pressure in the extra dimension is  allowed to be different from that of the pressure of the 3D space.  Within this framework, we arrive at an underdetermined system of coupled differential equations.   This is exactly like that of 4D FRW cosmology where the variables are left underdetermined by the field equations and one usually chooses an equation of state to proceed.   By adopting two equations of state, each relating the 3D and higher-dimensional pressures to the density, we obtain an exact expression relating the higher-dimensional scale factor to a function of the 3D scale factor and two arbitrary integration constants.  Surprisingly, when one of the integration constants is set equal to zero, a generalized case of the dynamical compactification scenario of Mohammedi \cite{Mohammedi} is found and represents a special case of the general treatment.  The aforementioned expression relating the higher-dimensional scale factor to the 3D scale factor allows us to write the 5D FRW field equations exclusively in terms of the 3D scale factor and we arrive at a set of decoupled 4D effective FRW field equations.

This paper is organized as follows.  In Sec. II, we subject the 5D Einstein field equations to a flat, anisotropic Robertson-Walker metric.  By adopting two equations of state, we arrive at an expression for the higher-dimensional scale factor as a function of the 3D scale factor.   In Sec. III, we write the FRW field equations exclusively in terms of the 3D scale factor and hence arrive at a set of 4D effective FRW field equations.  In Sec. IV, we find an exact expression relating a function of the 3D scale factor to the time.  We expand this exact expression in two different limiting regimes and obtain the approximate solutions corresponding to that of a generalized treatment of the case of dynamical compactification of Mohammedi \cite{Mohammedi} and to the 5D vacuum solution of Chodos and Detweiler \cite{Chodos}.  We then expand the hypergeometric function near a branch point and obtain a perturbative solution for the 3D scale factor in the small time regime.  We then show that this early-time solution exhibits accelerated expansion.   Finally, in Sec. V we summarize our results.

\section{5D anisotropic FRW cosmology \label{framesec}}

We begin with the Einstein field equations in five dimensions of the form 
\beq\label{field}
 G_A{}^B=\bar{\kappa} \;T_A{}^B,
  \eeq
where $A,B$ are indices which run over all spacetime dimensions.  In this manuscript, we set the higher-dimensional coupling constant equal to one, $\bar{\kappa}=1$.  

The higher-dimensional stress-energy tensor will be assumed to be that of a perfect fluid and of the form
\beq\label{stress}
T_A {}^ B= \textrm{diag} \left[ -\rho(t), p(t) , p(t), p(t), p_5(t)\right],
\eeq
where $p_5(t)$ is the pressure in the extra dimension.  As is obvious from (\ref{stress}), we are allowing the pressure in the extra dimension to be different, in general, from the pressure in the 3D space.  Hence, this stress-energy tensor describes a homogeneous, anisotropic perfect fluid in five dimensions.

We choose a metric ansatz of the form
\beq\label{metric} 
\dd s^2 = - dt^2 + a^2(t)\left[ dr^2 + r^2 \left(d\theta^2 + \sin^2\theta d\phi^2 \right) \right] + b^2(t)\;dy^2
\eeq
where we allow the scale factor of the extra dimension, $b(t)$, to evolve at a different rate, in general, than the 3D scale factor, $a(t)$.   This metric ansatz describes a flat, homogeneous and isotropic 3D space and a flat extra dimension.  Notice that (\ref{stress}) and (\ref{metric}) encompass the main assumptions of this manuscript.

By adopting the above metric ansatz and the perfect fluid stress-energy tensor, the 5D Friedmann-Robertson-Walker (FRW) field equations and the 5D conservation equation are of the form
\bea
\rho&=&3\frac{\dot a^2}{a^2} +3\frac{\dot a\dot b}{ab}\label{density}\\
p&=&- \left[2 \frac{\ddot a}{a} +\frac{\dot a^2}{a^2}+\frac{\ddot b}{b}+2\frac{\dot a\dot b}{ab}\right] \label{pres}  \\
p_5&=&- 3\left( \frac{\ddot a}{a} +\frac{\dot a^2}{a^2}\right)\label{presD}\\
0&=&\dot{\rho}+3\frac{\dot{a}}{a}(\rho+p)+\frac{\dot{b}}{b}(\rho+p_5),\label{cons}
\eea
where a dot denotes a time derivative.  One can easily show that the conservation equation, (\ref{cons}), is in fact satisfied when (\ref{density}) - (\ref{presD}) are employed.
Hence, the physical variables of the field equations, for general density and pressures, are left underdetermined as we have three unique equations and five unknowns. This is analogous to standard 4D FRW cosmology where one usually adopts an EoS relating the density to the pressure to proceed.  Here, we adopt two equations of state of the form
\bea
p&=&w\;\rho\label{EoS}\\
p_5&=&v\;\rho,\label{EoSd}
\eea
thus relating the 3D and higher-dimensional pressures to the density.  Notice that the EoS parameters $w$ and $v$ can in general be time-dependent, in this manuscript they are simply treated as constants.  

Employing (\ref{EoS}) and (\ref{EoSd}) to eliminate $\rho$, $p$, and $p_5$ from (\ref{density}) - (\ref{presD}) and performing some algebra, we obtain an exact differential equation of the form
\beq\label{diff}
\frac{\dd}{\dd t}\left[a^{3-n}\;\frac{\dd}{\dd t}\left(a^nb\right)\right]=0,
\eeq
where we defined the parameter
\beq\label{n} 
n\equiv\left[\frac{3w-2v-1}{1-v}\right].
\eeq
Notice that when the EoS parameters $w$ and $v$ are set equal to each other, the above parameter is negative and takes on the value $n=-1$.  

Equation (\ref{diff}) can be integrated twice, and remarkably, the scale factors can be decoupled.  Performing the aforementioned integrations, one finds a relation for the higher-dimensional scale factor in terms of the 3D scale factor of the form

\beq\label{bfuna}
b(t)= \frac{1}{a^{n}(t)}\left[\gamma_1+\gamma_0 \int a(t)^{(n-3)}dt\right],
\eeq
where $\gamma_0$ and $\gamma_1$ are arbitrary constants of integration.  

It should be emphasized that (\ref{bfuna}) is an exact expression.  Notice that if the arbitrary integration constant $\gamma_0$ is set equal to zero, we obtain the dynamical compactification scenario of Mohammedi \cite{Mohammedi} where, there, $n$ was an arbitrary power that was held positive.  Here, dynamical compactification emerges naturally as a special case of an exact treatment with the value of $n$ determined by the EoS parameters $w$ and $v$.  The discovery of (\ref{bfuna}) allows us to decouple the 5D FRW equations and obtain a set of differential equations written exclusively in terms of the 3D scale factor.  In the next section, we arrive at this set of 4D effective FRW field equations. 

\section{4D Effective FRW Field Equations}
Now that we have an expression for the higher-dimensional scale factor $b(t)$ in terms of the 3D scale factor $a(t)$, we can write the FRW field equations exclusively in terms of the 3D scale factor.  After performing some algebra, we find that (\ref{density}) - (\ref{cons}) can be written in the form
\bea
\rho&=&\eta_1\frac{\dot a^2}{a^2}+3\frac{\gamma_0}{x}\frac{\dot a}{a}\label{rho0}\\
 p&=&\eta_2\frac{\ddot a}{a}+(\eta_1-\eta_2^2)\frac{\dot a^2}{a^2}-(\eta_1+2\eta_2)\frac{\gamma_0}{x}\frac{\dot a}{a}\label{p0}\\
 p_5&=&-3\left(\frac{\ddot a}{a}+\frac{\dot a^2}{a^2}\right)\label{pd0}\\
0&=&\frac{\dot{\rho}}{\rho}+3(1+w-\frac{1}{3}n(1+v))\frac{\dot{a}}{a}+(1+v)\frac{\gamma_0}{x}\;,\label{cons0}
\eea
where we defined the coefficients
\bea
\eta_1&\equiv&3(1-n)\label{eta1}\\
\eta_2&\equiv&(n-2)\label{eta2}\;,
\eea
where $n$ was defined in (\ref{n}).  We also defined the higher-dimensional volume element, $x$, as 
\beq\label{volume}
 x(t)\equiv a^3b=a(t)^{(3-n)}\left[\gamma_1+\gamma_0\int a(t)^{(n-3)}dt\right],
\eeq
where we used (\ref{bfuna}) to express the volume element solely in terms of $a(t)$.
As a check of the 4D effective field equations, one can solve (\ref{rho0}) and (\ref{p0}) for $\dot a^2/a^2$ and $\ddot a/a$ in terms of the density, pressure, and the higher-dimensional volume element and then substitute these expressions into (\ref{pd0}).  Using (\ref{EoS}) and (\ref{EoSd}) and performing some algebra, one can in fact verify that (\ref{pd0}) is a redundant equation, as should be expected.

As in 4D FRW cosmology, the 5D conservation equation yields an expression for the density as a function of the 3D scale factor.  Integrating (\ref{cons}) and then employing (\ref{bfuna}) and (\ref{volume}), we find an expression for the density of the form
\beq\label{rhofuna}
\rho(a)=\rho_0\frac{a^{3(v-w)}}{x^{(1+v)}},
\eeq
where $\rho_0$ is a constant of integration.  Notice that in order to obtain a constant energy density, we must set $w=v=-1$, as is evident from (\ref{rhofuna}).  Also notice that when the higher-dimensional EoS parameter, $v$, is set to $v=-1$, (\ref{cons0}) and (\ref{rhofuna}) reduce to that of a 4D form.  This special case equates to the first law of thermodynamics for cosmology in 4D.  Hence, when $v=-1$, the dynamical evolution of the density is independent of the extra dimension.

In obtaining the 4D effective field equations of the form (\ref{rho0}) - (\ref{pd0}), we used the fact that
\beq\label{hubble}
\frac{\dot b}{b}=-n\frac{\dot a}{a}+\frac{\gamma_0}{x}.
\eeq
This expression relates the higher-dimensional Hubble parameter to the 3D Hubble parameter and the higher-dimensional volume element.  Notice that the higher-dimensional Hubble parameter can change signs with the sign dependent on whether the 4D Hubble parameter or the inverse of the higher-dimensional volume element dominates, assuming $\gamma_0$ is positive.  

By defining an \textit{effective} EoS parameter $\tilde{w}$ to be of the form
\beq\label{effw}
\tilde{w}\equiv w-\frac{1}{3}n(1+v),
\eeq
the 5D conservation equation, (\ref{cons0}), reduces to a form that precisely mimics the 4D conservation equation when the integration constant $\gamma_0$ is set equal to zero, with $w$ replaced by $\tilde{w}$.  This effective EoS parameter equates to an effective pressure of the form
\beq\label{effective}
\tilde{p}\equiv p-\frac{1}{3}n(\rho+p_5)=\tilde{w}\rho.
\eeq
This effective pressure was originally defined in \cite{Mohammedi} and proves useful in the following sections.  By employing (\ref{effective}), the 4D effective FRW equations take the form
\bea
\rho&=&\eta_1\frac{\dot a^2}{a^2}+3\frac{\gamma_0}{x}\frac{\dot a}{a}\label{rho2}\\
\tilde{p}&=&-\frac{1}{3}\eta_1\left(2\frac{\ddot a}{a}+\frac{\dot a^2}{a^2}\right)+\frac{\gamma_0}{x}\frac{\dot a}{a}\label{p2}\\
p_5&=&- 3\left( \frac{\ddot a}{a} +\frac{\dot a^2}{a^2}\right),\label{pd2}
\eea
where we have included (\ref{pd2}) for completeness, although it represents a redundant equation.

Before proceeding to a general treatment of the 4D effective field equations, we first discuss the conditions that are necessary in order to obtain accelerated expansion.  Combining (\ref{rho2}) and (\ref{p2}), we obtain an expression for the acceleration of the form
\beq\label{acceleration}
\frac{\ddot a}{a}=\frac{3}{\eta_1}\left[-\frac{1}{6}(1+3\tilde{w})\rho+\frac{\gamma_0}{x}\frac{\dot a}{a}\right].
\eeq
As is obvious from (\ref{acceleration}), the value of the acceleration is determined by two competing terms.  When the first term on the right-hand side dominates, one has accelerated expansion when $\tilde{w}<-1/3\;\mbox{and}\;\eta_1>0$, which can potentially occur for positive values of the EoS parameters $w$ and $v$.  This was first pointed out in \cite{Mohammedi} where, there, $n$ was an arbitrary parameter that was held positive to ensure dynamical compactification.  In the present treatment, $n$ is no longer an arbitrary power but rather determined by the EoS parameters $w$ and $v$ (see (\ref{n})).  When the second term on the right-hand side dominates, one obtains accelerated expansion when $\eta_1\;\mbox{and}\;\gamma_0$ are of the same sign.  

These two scenarios emerge as limiting cases of a general treatment and will be discussed thoroughly in the appendix.  In the next section, we proceed with a general treatment of the 4D FRW field equations.

\section{General Treatment of the 4D Effective Field Equations}\label{GT}

Using (\ref{EoSd}) to eliminate $\rho$ and $p_d$ from (\ref{rho2}) and (\ref{pd2}) and rearranging, we obtain an expression dictating the behavior of the 3D scale factor of the form
\beq\label{gt}
\gamma_1+\gamma_0\int a^{n-3}\;dt=-v\gamma_0\cdot a^{n-3}\left[\frac{\frac{\dd}{\dd t}\;a^{3(1+\tilde{w})/2}}{\frac{\dd^2}{\dd t^2}\;a^{3(1+\tilde{w})/2}}\right]\;\;\;\;\mbox{for}\;\;\;\; v\neq 0\;, \tilde{w}\neq -1, 
\eeq
where we used the fact that 
\beq\label{etapos}
\eta_1=\frac{3}{2v}(3\tilde{w}-1).
\eeq
The special cases of $v=0$ and $\tilde{w}=-1$ will be treated separately in Appendixes \ref{v0} and \ref{w-1}, respectively, where the exact analytical expressions for the 3D scale factor are presented.  Notice that (\ref{gt}) can equivalently be obtained from (\ref{rho2}) and (\ref{p2}) by using (\ref{effective}) to eliminate $\rho$ and $\tilde{p}$ from the field equations, as should be expected.

Taking a derivative of (\ref{gt}) to eliminate the integral, rearranging, integrating, and performing some algebra, we obtain an expression of the form
\beq\label{int2}
\dot{g} g^{2(n-3)/3(1+\tilde{w})}=c_0\;\ddot g\;\dot g^{-1/v}\;\;\;\;\mbox{for}\;\;\;\; v\neq 0\;, \tilde{w}\neq -1\;,
\eeq
where the constant of integration is found to be positive, $c_0\geq0$, and we defined the function 
\beq\label{ga}
g\equiv a^{3(1+\tilde{w})/2}.
\eeq

Equation (\ref{int2}) can be integrated with the left- and right-hand sides of the equation, in general, giving rise to powers of $g$ and $\dot g$, respectively.  There exists, however,  two special cases of EoS parameters when the integration of (\ref{int2}) yields logarithmic expressions of $g$ and $\dot{g}$.\footnote{The special case of $v=1$ will not be addressed in this manuscript.}  Notice that the effective EoS parameter $\tilde{w}=1/3$ when $\eta_1=0$ and the subsequent integration of the left-hand side yields a logarithmic expression for $g$.  This special case, which dramatically simplifies the field equations (\ref{rho2}) and (\ref{p2}), will be addressed in appendix \ref{eta0}.

Now integrating (\ref{int2}), rearranging, and again integrating, we find an exact expression relating the 3D scale factor to the time of the form 
\bea\label{hyper}
a^{3(1+\tilde{w})/2}&\cdot&_2F_1\left(\frac{v}{v-1}\;,\; \frac{3(1+\tilde{w})}{(3\tilde{w}-1)}\cdot\frac{v}{v-1}\; ;\; 1+\frac{3(1+\tilde{w})}{(3\tilde{w}-1)}\cdot\frac{v}{v-1} \;;\;z\;\right)\nonumber\\
&=&\left[\frac{(1-3\tilde{w})}{3(1+\tilde{w})}\;\frac{c_0}{c_2}\right]^{v/(1-v)} (t-t_0)\;\;\;\;\;\;\;\;\mbox{where}\;\;\;\;\;\;\;\;z\equiv\frac{1}{c_2a^{\eta_1(1-v)/3}}\;\;,
\eea
where $c_2$ and $t_0$ are constants of integration and we used (\ref{etapos}) and (\ref{ga}) to obtain this result.  Notice that ${}_2F_1(\alpha, \beta ; \gamma ; z)$ is a hypergeometric function that can be defined in terms of a hypergeometric series of the form
\beq\label{series}
_2F_1(\alpha,\beta; \gamma; z)=1+\frac{\alpha\beta}{\gamma}\;z+\frac{\alpha(\alpha+1)\beta(\beta+1)}{2\gamma(\gamma+1)}\;z^2+...
\eeq
In general, the hypergeometric series has singularities at $z=0,\;1,\mbox{and}\;\infty$, with a branch point at $z=1$, and converges absolutely throughout the entire unit circle $|z|<1$ in the complex plane if $\mbox{Re}\;(\alpha+\beta-\gamma)<0$ (see  \cite{table, table2}).  Absolute convergence  of the hypergeometric series of (\ref{hyper}) equates to
\beq\label{ineq}
a(t)>c_2^{-3/\eta_1(1-v)}\;\;\;\;\mbox{when}\;\;\;\;\frac{1}{1-v}>0.
\eeq
If one demands that the 3D scale factor be \textit{real} and positive for all values of the EoS parameters,  then the integration constant $c_2\geq0$.  Notice that the second inequality in (\ref{ineq}) is always satisfied when the higher-dimensional EoS parameter lies within the range $-1\leq v<1$.

\subsubsection{$z\ll1$}

Writing the hypergeometric function in (\ref{hyper}) as a hypergeometric series is useful for obtaining the approximate behavior of (\ref{hyper}) in the small $z$ limit.  Applying the series expansion (\ref{series}) to (\ref{hyper}) and keeping only the lowest-order contribution, the 3D scale factor takes the form 
\beq\label{asol2}
a(t)\simeq\left[\frac{(1-3\tilde{w})}{3(1+\tilde{w})}\frac{c_0}{c_2}\right]^{2v/3(1+\tilde{w})(1-v)}t^{2/3(1+\tilde{w})}\;,
\eeq
where, without loss of generality, we set the integration constant $t_0$ equal to zero.  This solution for the 3D scale factor has the same functional form as that of the approximate solution of Appendix \ref{ACII} and as that of standard 4D FRW cosmology with $w$ replaced with $\tilde{w}$, where $\tilde{w}$ is defined in (\ref{effw}).   In Appendix \ref{ACII}, we find that the approximate solution presented there is subjected to the parameter constraint $\eta_1>0$ and is valid in the late universe when $t$ is large.   As small $z$ corresponds to a large 3D scale factor for positive $\eta_1$ [see the definition of $z$ in (\ref{hyper})], we find that (\ref{asol2}) and (\ref{sol1}) agree in their regimes of validity and represent the same approximate solution.

Given that our constants of integration $c_0$ and $c_2$ are both positive, we find the additional parameter constraints
\beq\label{req}
\tilde{w}< \frac{1}{3}\;\;,\;\;v< 0
\eeq
when the approximate solution of (\ref{asol2}) is valid.  The parameter constraints of (\ref{req}) emerge if one demands that the 3D scale factor is real and we employ (\ref{etapos}).

The approximate solution of (\ref{asol2}) corresponds to that of a generalized treatment of the case of {\it dynamical compactification} of \cite{Mohammedi,others, Pan}.  In \cite{Mohammedi}, the higher-dimensional scale factor was {\it assumed} to dynamically evolve through a relation of the form 
\beq\label{b}
b(t)\sim\frac{1}{a^n(t)}
\eeq
where $n$, there, was an arbitrary parameter constrained only by the relation $n> 0$.   By hypothesizing (\ref{b}), it was shown in \cite{Mohammedi} that the higher-dimensional FRW field equations reduce precisely to that of a 4D form, once an effective pressure of the form (\ref{effective}) is adopted.  Further, it was realized that the effective EoS parameter $\tilde{w}$ could in fact be negative for positive values of $w$ and $v$.  Thus, one could have accelerated expansion of the 3D spatial manifold for positive EoS parameters $w$ and $v$.

In this manuscript, $n$ is no longer an arbitrary power but rather a function of the EoS parameters, as defined in (\ref{n}).   When the solution of this subsection is valid, we find that the higher-dimensional scale factor does in fact take the form of (\ref{b}), but emerges as a limiting case of the exact expression of (\ref{bfuna}).  For this generalized treatment of dynamical compactification, which is exhaustively studied and presented in Appendix \ref{ACII}, we find that one cannot obtain accelerated expansion of the 3D scale factor in the late universe for positive values of density and pressure.\footnote{see Fig. \ref{fig:AccelExpnPos} in Appendix \ref{ACII} for a plot of the EoS parameter space that does yield accelerated expansion.}  In addition, we find that if the 3D scale factor is in fact undergoing accelerated expansion in the late universe, then the higher-dimensional scale factor cannot be dynamically compactifying.


\subsubsection{$z\gg1$}

The large $z$ behavior of  (\ref{hyper}) can also be investigated by first applying a hypergeometric transformation formula \cite{table} of the form
\bea\label{trans1}
_2F_1(\alpha,\beta; \gamma ;z)&=&\frac{\Gamma(\gamma)\Gamma(\beta-\alpha)}{\Gamma(\beta)\Gamma(\gamma-\alpha)}\;(-z)^{-\alpha}\;_2F_1(\alpha\;,\; 1-\gamma+\alpha \;;\;1-\beta+\alpha\;;\; 1/z)\nonumber\\
&+&\frac{\Gamma(\gamma)\Gamma(\alpha-\beta)}{\Gamma(\alpha)\Gamma(\gamma-\beta)}\;(-z)^{-\beta}\;_2F_1(\beta\;,\;1-\gamma+\beta\; ;\; 1-\alpha+\beta\;;\; 1/z)
\eea
to (\ref{hyper}) and then by expanding as a hypergeometric series for small $1/z$.  Employing (\ref{trans1}) and  applying the series expansion (\ref{series}) for small $1/z$, the 3D scale factor takes the form
\beq\label{abvac2}
a(t)\simeq\left[\frac{4}{3(1+\tilde{w})}\left[\frac{(3\tilde{w}-1)}{3(1+\tilde{w})}c_0\right]^{v/(1-v)}\right]^{1/2}t^{1/2}\;,
\eeq 
where we kept only the lowest-order contribution. 
It should be noted that when the hypergeometric transformation (\ref{trans1}) is applied to (\ref{hyper}), the second term of the transformation equation reduces to merely a function of the EoS parameters and the integration constant $c_2$.   This term, which is void of the 3D scale factor, can be absorbed into the integration constant $t_0$, which effectively shifts the initial value of the time.   

The approximate solution of (\ref{abvac2}) has precisely the same functional form as that of the approximate solution of Appendix \ref{ACI}.  The approximate solution of Appendix \ref{ACI} is found to be valid in the late universe when $t$ is large and when $\eta_1<0$.  As small $1/z$ corresponds to a large 3D scale factor for negative $\eta_1$ [again see the definition of $z$ in (\ref{hyper})], we find that (\ref{abvac2}) and (\ref{abvac}) agree in their regimes of validity and represent the same approximate solution.

As the constant of integration $c_0\geq0$, we find the parameter constraints
\beq\label{req2}
\tilde{w}> \frac{1}{3}\;\;,\;\;v< 0
\eeq
when the approximate solution of (\ref{abvac2}) is valid.  The parameter constraints of (\ref{req2}) emerge if one again demands that the 3D scale factor is real and we employ (\ref{etapos}).

\subsubsection{$z\sim 1$}\label{z1}
We now explore the behavior of (\ref{hyper}) in the vicinity of the $z=1$ branch point.  This equates physically to an expansion of the 3D scale factor about its minimum value, which will become obvious later in this subsection.  

To explore the behavior of the 3D scale factor in this regime, we employ another hypergeometric transformation formula \cite{table} of the form
\bea\label{trans}
_2F_1(\alpha,\beta; \gamma ;z)&=&\frac{\Gamma(\gamma)\Gamma(\alpha+\beta-\gamma)}{\Gamma(\alpha)\Gamma(\beta)}\;(1-z)^{\gamma-\alpha-\beta}\;_2F_1(\gamma-\alpha, \gamma-\beta ;\gamma-\alpha-\beta+1;1-z)\nonumber\\
&+&\frac{\Gamma(\gamma)\Gamma(\gamma-\alpha-\beta)}{\Gamma(\gamma-\alpha)\Gamma(\gamma-\beta)}\;_2F_1(\alpha,\beta ; \alpha+\beta-\gamma+1;1-z).
\eea
Applying this transformation formula to the hypergeometric function in (\ref{hyper}) and performing some algebra, one can show that (\ref{hyper}) takes the form
\bea\label{hyper1}
&&a^{3(1+\tilde{w})/2}\left[1-\frac{1}{c_2a^{\eta_1(1-v)/3}}\right]^{1/(1-v)}\nonumber\\
&\cdot&_2F_1\left(\frac{1}{1-v}\left[1+\frac{3(1+\tilde{w})}{(1-3\tilde{w})}v\right],1;\frac{2-v}{1-v};\;1-\frac{1}{c_2a^{\eta_1(1-v)/3}}\right)=-\frac{1}{v}\left[\frac{(1-3\tilde{w})}{3(1+\tilde{w})}\frac{c_0^v}{c_2^v}\right]^{1/(1-v)}t\nonumber\\
\eea
where, again, $t$ represents a shifted time coordinate as the second term of the transformation equation is void of the 3D scale factor and can be absorbed into the integration constant $t_0$.  

We now employ the series expansion (\ref{series}) for small $(1-z)$ to the hypergeometric function in (\ref{hyper1}), perform some algebra, and obtain an expression of the form
\beq\label{f4}
\frac{(1-v)}{(2-v)}\left[1-\frac{3(1+\tilde{w})}{(1-3\tilde{w})}\cdot\frac{v}{1-v}\right]f^2+f+\frac{(1-3\tilde{w})}{2}\frac{(1-v)}{v}\kappa\;t^{(1-v)}=0\;,
\eeq
where we kept terms up to order $o(f^2)$ in (\ref{f4}) and defined the quantities
\bea\label{kappa}
f&\equiv&1-\frac{1}{c_2a^{\eta_1(1-v)/3}}\label{f1}\\
\kappa&\equiv& \left[\frac{2}{3(1+\tilde{w})}\cdot\frac{(-v)^v}{(1-v)}\right]\frac{c_0^v}{c_2^{4v/(1-3\tilde{w})}}\label{kappa}.
\eea

As we are interested in the time evolution of the 3D scale factor for small $f$, (\ref{f4}) can be solved perturbatively by writing $f$ as a series solution of the form
\beq
f=f_0+f_1+...
\eeq
where $f_0$ is the solution to the linear approximation of (\ref{f4}), $f_1$ is the first-order correction term, etc.  Using this method and the definition of $f$ in (\ref{f1}), we obtain a perturbative solution for the 3D scale factor of the form
\beq\label{asol}
a(t)=c_2^{-3/\eta_1(1-v)}\left[1+\kappa t^{(1-v)}+\frac{(1-v)}{4(2-v)}\left[(1-3\tilde{w})-\frac{2(2-3v)}{(1-v)}\right]\kappa^2 t^{2(1-v)}+o(\kappa^3 t^{3(1-v)})\right],
\eeq
where we expanded the above expression for the scale factor for small $\kappa t^{(1-v)}$ and kept terms up to order $o(\kappa^2t^{2(1-v)})$.   Notice that $t$  is the time interval from when $a(t)|_{t=0}=c_2^{-3/\eta_1(1-v)}$.

If one demands that the 3D scale factor is real and positive for all values of the EoS parameters, then in addition to the requirement that $c_2\geq0$, we also find the additional constraint 
\beq
v\leq 0,
\eeq
where we used our definition of $\kappa$ in (\ref{kappa}) to obtain this result.  Remarkably, when the higher-dimensional EoS parameter, $v$, is in fact negative, we find that the 3D scale factor exhibits {\it accelerated expansion} for small time.  For the special cases of $w=v/3$ and of $w=0\;,\,v=-1$, the exact analytical solutions for the 3D scale factor are found and presented in Appendixes \ref{wv/3} and \ref{Aw0v-1}, respectively.  For these special cases the 3D scale factor exhibits an epoch of early-time accelerated expansion, which abruptly ends when an inequality involving the time ceases to be satisfied.  Hence, we see that this model of anisotropic evolution has a natural mechanism for turning off an early-time inflationary epoch.

Inserting the solution for the 3D scale factor, (\ref{asol}), back into the 4D effective FRW field equations (\ref{rho2}) - (\ref{pd2}), we find a consistent solution up to and including order $o(\kappa^2t^{2(1-v)})$ when $\gamma_1$ is set equal to zero and 
\beq\label{constant2}
\frac{\rho_0}{\gamma_0^{(1+v)}}=\frac{2}{(1+\tilde{w})}\cdot c_2(-vc_0)^v.
\eeq	
Now, inserting (\ref{asol}) into (\ref{bfuna}) and setting $\gamma_1=0$, the higher-dimensional scale factor takes the form
\beq\label{bsolgen}
b(t)=\gamma_0\;c_2^{9/\eta_1(1-v)}\cdot t\left[1-\frac{(1-v)}{2v(2-v)}\left[(1-3\tilde{w})+\frac{2v(4-v)}{(1-v)}\right]\kappa t^{(1-v)}+o(\kappa^2t^{2(1-v)})\right]
\eeq
for small $\kappa t^{(1-v)}$, where we kept terms up to and including order $o(\kappa t^{(1-v)})$.  Notice that the higher-dimensional scale factor goes to zero and the density goes to infinity as $t\rightarrow 0$.  This behavior of the higher-dimensional scale factor differs dramatically from that of the 3D scale factor, which approaches a constant for vanishing time.  Also notice that the higher-dimensional scale factor expands for small $t$.

\section{Conclusion} \label{conclusion}

In conclusion, we examined the time evolution of the five-dimensional Einstein field equations subjected to a flat, anisotropic Robertson-Walker metric where the scale factors for the 3D and higher-dimensional spatial manifolds are allowed to evolve independently of each other and at different rates.  We chose the stress-energy tensor to be that of a perfect fluid where we allowed the 3D and higher-dimensional pressures to, in general, be different.  By adopting equations of state relating the 3D and higher-dimensional pressures to the density,  we obtained an exact expression relating the higher-dimensional scale factor to a function of the 3D scale factor.  This expression allowed us to eliminate the higher-dimensional scale factor from the FRW field equations and to arrive at a set of 4D effective FRW field equations, written exclusively in terms of the 3D scale factor.

We examined the 4D effective FRW field equations in the general case and obtained an exact expression that relates a function of the 3D scale factor to the time.  This exact expression includes a power of the 3D scale factor and a hypergeometric function and cannot, in general, be inverted to yield an analytical expression for the 3D scale factor as a function of the time.   By expanding the hypergeometric function for small argument, we obtained an approximate solution that corresponds to that of a generalized treatment of the case of dynamical compactification of Mohammedi \cite{Mohammedi}.   In \cite{Mohammedi}, it was assumed that the higher-dimensional scale factor dynamically compactified as the 3D scale factor expanded through a power law relation.  By assuming this functional form of dynamical compactification, it was found in \cite{Mohammedi} that the effective pressure can in fact be negative for positive values of the 3D and higher-dimensional pressures, hence allowing for accelerated expansion with positive pressures.  In this manuscript it is found that the exact functional form of the dynamical compactification scenario of \cite{Mohammedi} emerges naturally in this limiting regime, where the parameter $n$  is no longer an arbitrary power but rather a function of the EoS parameters.  In the regime when this approximate solution is valid, we found that one can obtain accelerated expansion for the 3D scale factor only when the EoS parameters $w, v$ are both \textit{negative}.  Thus, one cannot obtain accelerated expansion for the 3D scale factor in the late universe for positive values of density and pressures.  In addition, we also found that if the 3D scale factor is in fact undergoing accelerated expansion in the late universe, the higher-dimensional scale factor cannot be undergoing dynamical compactification.  By then expanding the hypergeometric function for large argument, we obtained an approximate solution corresponding to that of the 5D vacuum solution of Chodos and Detweiler \cite{Chodos}.

We then explored the behavior of the hypergeometric function in the vicinity of $z=1$ and obtained the perturbative solution for the 3D scale factor in this regime.  We showed that by demanding that the 3D scale factor be \textit{real}, the higher-dimensional EoS parameter, $v$, must be \textit{negative} in this regime.  In addition, when the higher-dimensional EoS parameter is in fact negative in this regime, the 3D scale factor exhibits \textit{accelerated expansion}, independent of the value of the 4D equation of state parameter $w$.  We also showed that the higher-dimensional scale factor expands, to lowest order, in this regime.

It is interesting to compare the results of this manuscript to that of 4D FRW cosmology.  Here, we found that the 3D scale factor exhibits an epoch of accelerated expansion for small time, without the use of a scalar field.  Thus, this model offers a possible alternative to that of scalar field inflationary theory, although it is  thus far unclear if this model is capable of generating the irregularities necessary to lead to the formation of structure \cite{Liddle}.  The inflationary epoch of this model arises naturally from higher-dimensional geometry, so long as the higher-dimensional EoS parameter is negative.  In addition to this contrast with 4D FRW cosmology, we also note that when the time coordinate becomes vanishingly small, the 3D scale factor takes on a finite and nonzero value.  This differs sharply from that of 4D FRW cosmology where the 3D scale factor, in general, approaches zero for vanishing time.

The natural question that arises from this work concerns the physics of anisotropic evolution in $D$ dimensions.   We have found elsewhere that when one does in fact generalize to $D$ dimensions, the decoupling of the field equations arises in a similar fashion to that of the work presented here.  In addition, it is found that the 4D effective FRW field equations that emerge take on a similar form to those of  (\ref{rho2}) - (\ref{pd2}), with the presence of an additional term that vanishes uniquely in 5D.  This additional term complicates the field equations and surely alters the physics, when compared to the 5D treatment.  This will be the subject of a later work.

\appendix

\section{Approximate case I}\label{ACII}
Here we analyze the 4D effective FRW field equations subject to the approximation
\beq
\rho\gg \frac{\gamma_0}{x}\frac{\dot a}{a}.\label{approx2}
\eeq
Under this approximation, the 4D effective FRW equations (\ref{rho2}) - (\ref{p2}) take the form of
\bea
\rho&=&\eta_1\frac{\dot a^2}{a^2}\label{rho1}\\
\tilde{p}&=&-\frac{1}{3}\eta_1\left(2\frac{\ddot a}{a}+\frac{\dot a^2}{a^2}\right).\label{p1}
\eea
Notice that if one redefines the coupling constant to absorb the parameter $\eta_1$, Eqs. (\ref{rho1}) and (\ref{p1}) become precisely that of 4D FRW cosmology.   This result was first realized by Mohammedi in \cite{Mohammedi}.
Also notice that if one demands a positive higher-dimensional density, then $\eta_1>0$, as is evident from (\ref{rho1}).  This requirement is equivalent to $n<1$, which can easily be seen from (\ref{eta1}).  This parameter restriction constrains the possible values of the EoS parameters $w$ and $v$ and will be discussed in further detail later in this appendix.

Rearranging (\ref{rho1}) and (\ref{p1}), one finds an expression for the acceleration of the form
\beq
\frac{\ddot a}{a}=-\frac{1}{2\eta_1}\;(1+3\tilde{w})\rho.\label{acc0}
\eeq
It should be noted that one finds accelerated expansion for $-1\leq\tilde{w}<-1/3$, which, can potentially occur for positive values of the EoS parameters $w$ and $v$ (see \cite{Mohammedi}).  
This possibility will also be analyzed later in this appendix. 
 
We now proceed by finding the approximate solution for the 3D scale factor.  Using (\ref{effective}) to eliminate $\rho$ and $\tilde{p}$ from (\ref{rho1}) and (\ref{p1}), one finds a solution for the 3D scale factor of the form
\bea
a(t)&\simeq&a_0\;t^{2/3(1+\tilde{w})}\;\;\;\mbox{when}\;\;\tilde{w}\neq -1\label{sol1}\\
&\simeq&a_0\;e^{Ht}\;\;\;\;\;\;\;\;\;\;\mbox{when}\;\;\tilde{w}= -1.
\eea
This solution for the 3D scale factor has the same functional form as that of standard 4D FRW cosmology with $w$ replaced with $\tilde{w}$.  Notice that this solution is consistent with (\ref{rhofuna}), (\ref{rho1}), and (\ref{p1}) only when the additional approximation
\beq
\gamma_1\gg\gamma_0\int a(t)^{(n-3)}dt \label{approx1}
\eeq
is imposed or if the integration constant $\gamma_0$ is set equal to zero.   Inserting (\ref{sol1}) into (\ref{rho1}) and (\ref{p1}) and employing (\ref{rhofuna}) subject to (\ref{approx1}), $a_0$ is found to take the form
\beq\label{a0}
a_0=\left[\frac{3}{2}(1+\tilde{w})\sqrt{\frac{\rho_0}{\eta_1\gamma_1^{(1+v)}}}\;\;\right]^{2/3(1+\tilde{w})}\;\;\mbox{when}\;\;\tilde{w}\neq -1.
\eeq
This relation agrees with the value one obtains from 4D FRW cosmology once the coupling constant is redefined, as discussed earlier.  

Under the approximation of (\ref{approx1}), the higher-dimensional scale factor takes the form
\beq
b(t)\simeq\frac{\gamma_1}{a^{n}(t)}\label{ba},
\eeq
where we used (\ref{bfuna}).  Notice that if $n>0$, the higher-dimensional scale factor contracts as the 3D scale factor expands and one obtains the dynamical compactification scenario of \cite{Mohammedi}.

Now that we have the solutions for the scale factors, we need to check the consistency of the dual approximations of this appendix.  Inserting (\ref{sol1}) into (\ref{approx2}) and (\ref{approx1}), we find that the aforementioned approximations are in fact valid when
\beq\label{regime}
\frac{3(1+\tilde{w})}{2\eta_1}\cdot\frac{\gamma_0}{ \gamma_1}\;a_0^{n-3}\ll\;t\;^{2\eta_1(1-v)/9(1+\tilde{w})},
\eeq
where we used (\ref{etapos}).   As $\eta_1$ is strictly positive\footnote{Negative higher-dimensional energy densities are not considered in this manuscript.} and $\tilde{w}\geq-1$ for an expanding 3D spatial manifold, one finds that the power in (\ref{regime}) is in fact positive for $v<1$.  Hence, the approximate solution of this subsection is valid in the late universe when $t$ is large.   Further, notice that (\ref{etapos}), when subjected to the requirement of a positive higher-dimensional energy density, implies that $v$ {\it must} be negative if one is to have accelerated expansion of the 3D spatial manifold in this regime.

The approximations of this appendix, which give rise to the solutions of (\ref{sol1}) and (\ref{ba}), yield a generalized treatment of the case of dynamical compactification, which has been previously studied elsewhere (see \cite{Mohammedi, others, Pan}).   There, the higher-dimensional scale factor was {\it assumed} to dynamically evolve through a relation of the form of (\ref{ba}), where $n$ was an arbitrary parameter constrained only by the relation $n> 0$.   By hypothesizing dynamical compactification of the form given by (\ref{ba}), it was shown in \cite{Mohammedi} that the higher-dimensional FRW field equations reduce precisely to a 4D form, once an effective pressure of the form (\ref{effective}) is adopted.  Further, it was realized that the effective EoS parameter $\tilde{w}$ could in fact be negative for positive values of $w$ and $v$.  Thus, one could in fact have accelerated expansion of the 3D spatial manifold for positive values of $w$ and $v$.

In this manuscript, $n$ is no longer an arbitrary power but rather a function of the EoS parameters\footnote{Throughout  this work, the values of the EoS parameters $w$ and $v$ are constrained to the range $-1\leq w,v< 1$},  as defined in (\ref{n}).  We now wish to explore the EoS parameter space subjected to the constraints 
\bea
&\eta_1&>0\;\;\;\;\;\;\;\;\;\mbox{for positive energy density,}\nonumber\\
-1/3>&\tilde{w}&\geq -1\;\;\;\;\;\;\mbox{for accelerated expansion of the 3D scale factor,}\nonumber\\
&n&>0\;\;\;\;\;\;\;\;\;\mbox{for dynamical compactification}.
\eea

\begin{figure}
\begin{center}
\includegraphics[scale=0.7]{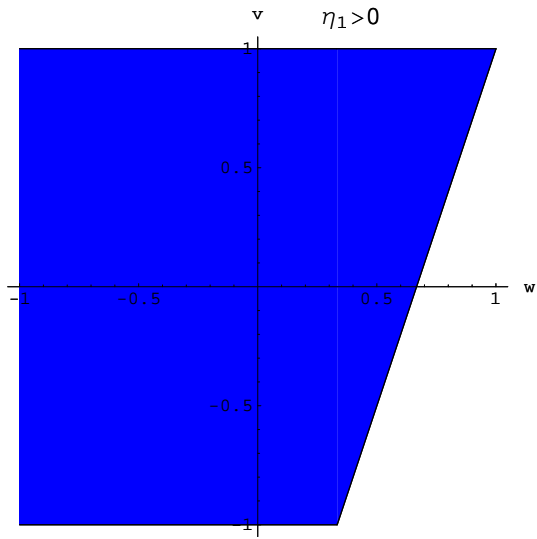}
\caption{EoS parameter space plot of $v$ versus $w$ subjected to $\eta_1>0$.  
}
\label{fig:eta1}
\end{center}
\end{figure}

The requirement that the higher-dimensional energy density remain positive is equivalent to the parameter constraint $\eta_1>0$.  Employing (\ref{eta1}) and (\ref{n}), this equates to an inequality of the form
\beq\label{posenergy}
v>3w-2
\eeq
for $v<1$.  Figure \ref{fig:eta1} shows a plot of the allowed parameter space subjected to the inequality of (\ref{posenergy}).  Notice that all allowed values of $v$ meet the requirement of $\eta_1>0$ when $w<1/3$.  The special case of $w=1/3$, $v=-1$, which yields $\eta_1=0$, will be further considered in Appendix \ref{eta0}, where an exact treatment is presented.

\begin{figure}
\begin{center}$
\begin{array}{cc}
\includegraphics[scale=0.7]{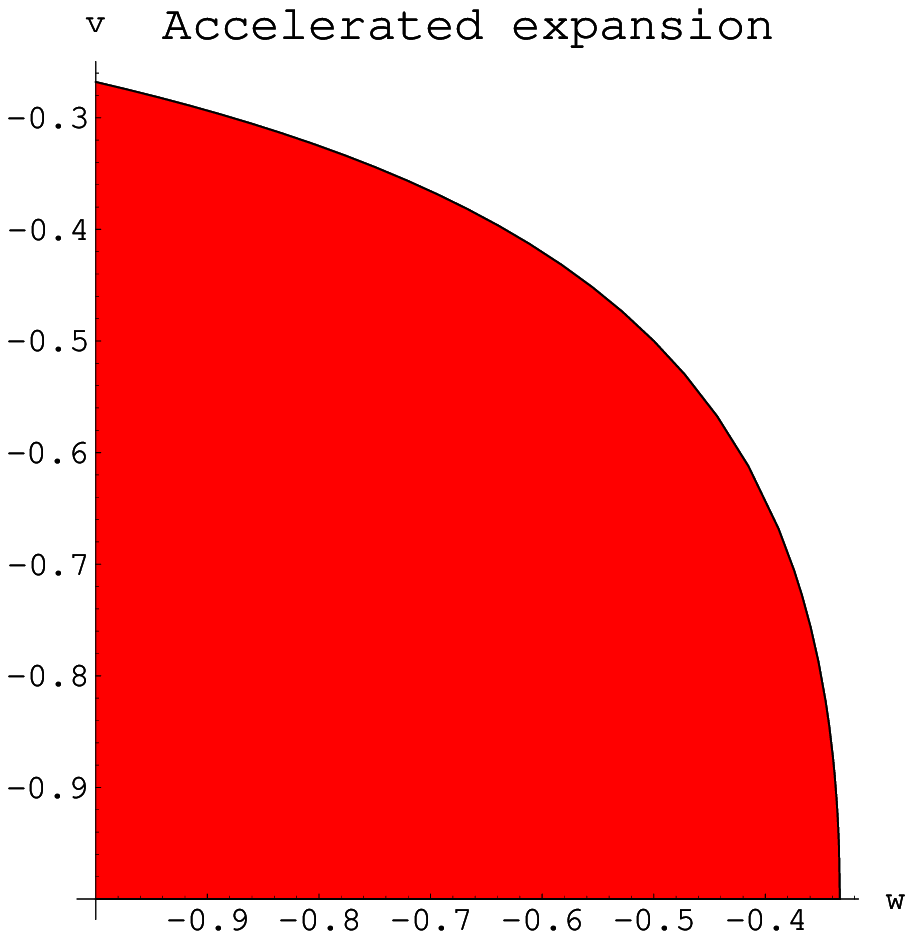}\;\;\;\;\;\;\;\;\;\;\;&\;\;\;\;\;\;\;\;\;\;\;
\includegraphics[scale=0.7]{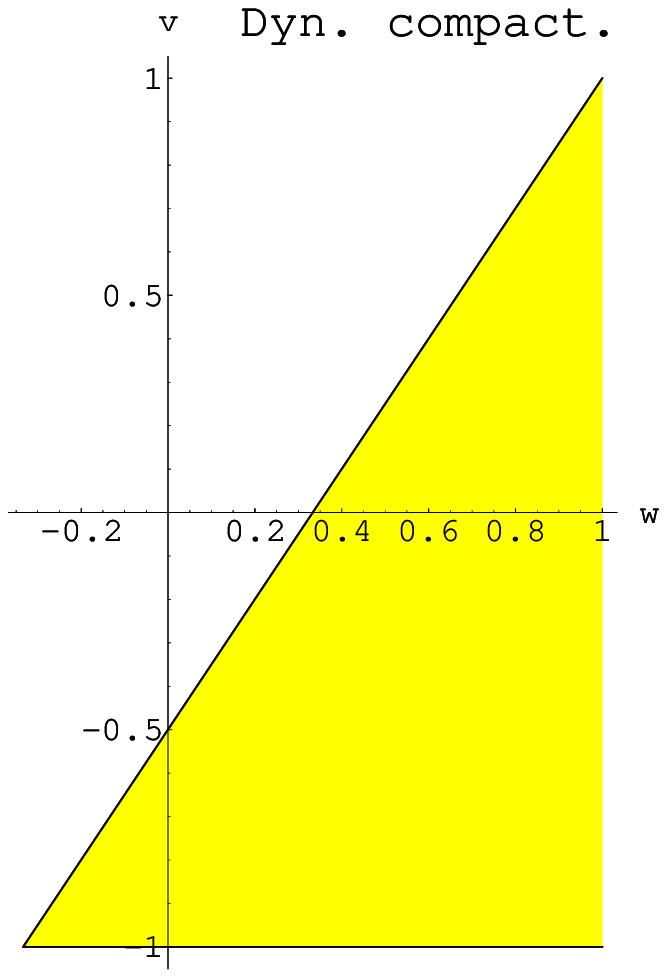}
\end{array}$
\caption{The plot on the left shows the EoS parameter space plot of $v$ versus $w$ subjected to $-1\leq\tilde{w}<-1/3$.  The plot on the right shows the EoS parameter space plot of $v$ versus $w$  subjected to $n>0$.  
}
\label{fig:AccelExpnPos}
\end{center}
\end{figure}

As the solution for the 3D scale factor found in this appendix is valid in the late universe, we  explore the parameter space when the EoS parameters yield accelerated expansion. This equates to the effective EoS parameter residing in the range $-1\leq\tilde{w}<-1/3$, as can be witnessed by (\ref{sol1}).   Employing (\ref{effw}) and (\ref{n}), this parameter constraint yields a set of inequalities, which are summarized by the expression
\beq\label{inequal}
\frac{v^2+2}{3v}\leq w<\frac{v^2+v+1}{3v}\;\;\mbox{for}\;v<0.
\eeq
It should be noted that for positive $v$ in the range $0<v<1$, there exist no values of $w$ that give rise to accelerated expansion of the 3D spatial manifold.  The left plot in Fig. \ref{fig:AccelExpnPos} shows the allowed parameter space subjected to the inequalities of (\ref{inequal}).  One finds accelerated expansion only when $w$ and $v$ are both negative and lie within the shaded region.  Upon comparison with Fig. \ref{fig:eta1}, one finds that this allowed parameter space, which yields accelerated expansion, also satisfies the requirement that $\eta_1>0$.    One cannot obtain accelerated expansion of the 3D scale factor in the late universe for positive values of the density and pressures when the approximations of this appendix are valid.  

Continuing, we now subject the EoS parameters to the requirement of dynamical compactification, which occurs when $n>0$.  Using (\ref{n}), this constraint yields an inequality of the form
\beq\label{ineq3}
v<\frac{1}{2}(3w-1).
\eeq
The right plot in Fig. \ref{fig:AccelExpnPos} shows the allowed parameter space subjected to (\ref{ineq3}).  Notice that when $w\leq1/3$, dynamical compactification only occurs for $v<0$.  Thus, if the 3D spatial manifold is dominated by pressureless matter or 4D radiation, one must have a negative higher-dimensional pressure for dynamical compactification to occur when the approximations of this appendix are valid.

\begin{figure}
\begin{center}
\includegraphics[scale=1.0]{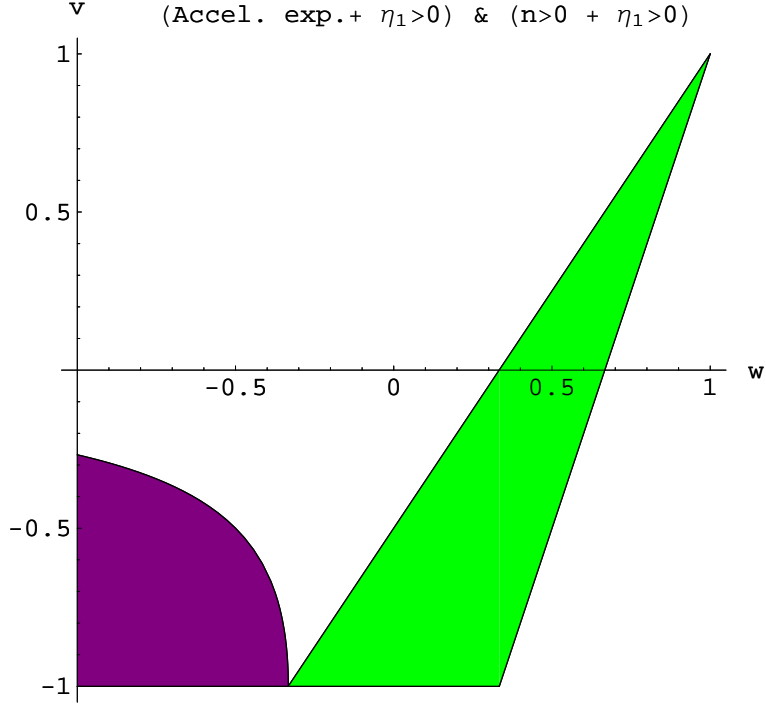} 
\caption{EoS parameter plots of $v$ versus $w$ for combinations of parameter constraints.  The region in the left of the plot shows the EoS parameter space subjected to accelerated expansion and a positive energy density.  The region in the right of the plot shows the EoS parameter space subjected to dynamical compactification and a positive energy density.}
\label{fig:triple}
\end{center}
\end{figure}

Finally, we explore the EoS parameter space subjected to all three of the aforementioned constraints.  Figure \ref{fig:triple} displays this parameter space and reveals two distinct regions of solutions.  The region in the lower left of the plot shows the allowed values of $w$ and $v$ that give rise to accelerated expansion and a positive higher-dimensional energy density.  The triangular wedgelike region to the right shows the allowed values of $w$ and $v$ subjected to the constraints of dynamical compactification and a positive higher-dimensional energy density.  Notice that these two superimposed plots yield \textit{no} shared region of EoS parameters.  Thus, one concludes that if the 3D scale factor is undergoing accelerated expansion in the late universe, when the approximations of this appendix are valid, then the higher-dimensional scale factor can {\it not} be dynamically compactifying.  A similar result was found in \cite{Pan}.

\section{Approximate case II}\label{ACI}
We now analyze the 4D effective FRW field equations subject to the approximation
\beq\label{approx3}
\rho\ll \frac{\gamma_0}{x}\frac{\dot a}{a}.
\eeq
For a positive 3D Hubble parameter, we find the constraint that $\gamma_0>0$ if the above approximation is to be valid as the density and higher-dimensional volume element are taken to be positive.

Under the approximation of (\ref{approx3}), the FRW equations (\ref{rho2}) - (\ref{p2}) take the form
\bea
0&=&\eta_1\frac{\dot a^2}{a^2}+3\frac{\gamma_0}{x}\frac{\dot a}{a}\label{rho3}\\
0&=&-\frac{1}{3}\eta_1\left(2\frac{\ddot a}{a}+\frac{\dot a^2}{a^2}\right)+\frac{\gamma_0}{x}\frac{\dot a}{a}.\label{p3}
\eea
These equations are easily solved by substitution.  We find an approximate solution for the 3D scale factor of the form
\beq\label{sol2}
a(t)\propto t^{1/2}.
\eeq
Inserting this solution for the 3D scale factor back into (\ref{rho3}) and (\ref{p3}), we find a consistent solution if a second approximation of the form
\beq
\gamma_1\ll\gamma_0\int a(t)^{(n-3)}dt \label{approx4}
\eeq
is made or, equivalently, if the integration constant $\gamma_1$ is set equal to zero.  Now, inserting (\ref{sol2}) into (\ref{bfuna}) subjected to the approximation of (\ref{approx4}), we obtain an approximate solution for the scale factors of the form
\bea\label{abvac}
a(t)&\simeq&a_0\;t^{1/2}\nonumber\\
b(t)&\simeq&b_0\;t^{-1/2}\;\;\mbox{where}\;\;b_0\equiv-\frac{6\gamma_0}{\eta_1a_0^3},
\eea
where $a_0$ is left undetermined by the field equations in this regime.
Notice that the time dependence of these solutions corresponds to that of the 5D vacuum solution of Chodos and Detweiler \cite{Chodos}, as one should expect.   If one demands that the scale factors remain positive, then $\gamma_0$ and $\eta_1$ must be of opposite signs.   As $\gamma_0>0$ for a positive 3D Hubble paramenter, we find that $\eta_1<0$.  Upon comparison of this sign requirement with the conditions needed for accelerated expansion, which was discussed in the paragraph following (\ref{acceleration}), we find a scenario which yields decelerated expansion.
This fact is explicitly demonstrated by the time dependence of the 3D scale factor in (\ref{abvac}).

Inserting (\ref{abvac}) into (\ref{approx3}) and (\ref{approx4}), we find a regime of validity given by the inequality
\bea
 \frac{\gamma_1}{b_0a_0^{n}}&\ll&t^{-\eta_1/6}.
\eea
Note that for $\eta_1<0$, we find that the treatment of this appendix is valid in the late universe.  This regime of validity matches that of the approximate solutions of Appendix \ref{ACII}; however, there the solutions were subject to the parameter constraint $\eta_1>0$.  Hence, the approximate solutions of Appendixes \ref{ACII} and \ref{ACI} represent different branches of approximate solutions in the late universe.

\section{special case of the general treatment: $\eta_1=0$}\label{eta0}

In this appendix, we address the special case of $\eta_1=0$, which corresponds to a relationship between the EoS parameters and to an effective EoS parameter of the form
\bea\label{specialcase}
v&=&3w-2\nonumber\\
\tilde{w}&=&1/3\;.
\eea
It is interesting to note that although $\tilde{w}=1/3$ when $v=0$ and when $v=3w-2$, the time evolution of the 3D scale factor is in fact quite different for each of these two possibilities.  

For the special choice of EoS parameters given by (\ref{specialcase}),  Eq. (\ref{int2}) can be written in terms of an exact differential equation of the form
\beq\label{specialint2}
\frac{\dd}{\dd t}\left[\ln g\right]=\frac{v}{(v-1)}c_0\frac{\dd}{\dd t}\left[\dot g^{(v-1)/v}\right].
\eeq
Now integrating (\ref{specialint2}), rearranging, and again integrating, we arrive at an exact expression relating the 3D scale factor to the time of the form
\beq\label{gamma}
\Gamma\left(\frac{1}{1-v}\;,\;z\right)=\frac{1}{\tilde{c}_1}\left[\frac{v}{1-v}c_0\right]^{v/(1-v)}(t-t_0)\;\;\;\;\;\;\;\;\mbox{where}\;\;\;\;z\equiv c_1-\ln g
\eeq
where $\Gamma(a,z)$ is the incomplete gamma function, $c_1$ and $t_0$ are integration constants, and $\tilde{c}_1\equiv e^{c_1}>0$.   It should be noted that (\ref{gamma}) is an exact expression relating the 3D scale factor to the time.  

The incomplete gamma function has a series representation given by
\beq\label{Gexp}
\Gamma\left(\alpha,\;z\right)=z^{\alpha}\left[-\frac{1}{\alpha}+\frac{z}{\alpha+1}-\frac{z^2}{2(\alpha+2)}+...\right]+\Gamma(\alpha)
\eeq
where $\Gamma(\alpha)$ is the gamma function.
We now proceed by applying this series representation to (\ref{gamma}) for small $z$ and keep only the first two terms of the series.  This small $z$ expansion equates physically to an expansion of the 3D scale factor about the value $\sqrt{\tilde{c}_1}$, which can be witnessed by examining our definition of $z$ and using (\ref{ga}) with $\tilde{w}=1/3$.   Applying this small $z$ expansion to the left-hand side of (\ref{gamma}) and performing some algebra, we obtain an expression of the form
\beq\label{z4}
-\frac{(1-v)}{(2-v)}z^2+z+\tilde{\kappa}\;t^{(1-v)}=0\;,
\eeq
where we kept terms up to order $o(z^2)$ in (\ref{z4}) and defined the parameter
\beq
\tilde{\kappa}\equiv\frac{(-v)^v}{2(1-v)}\frac{c_0^v}{\tilde{c}_1^{(1-v)}}\label{tildekappa}.
\eeq
Notice that we chose the value of the integration constant $t_0$ to cancel the gamma function term.  By making this choice, the time coordinate $t$ represents the time interval from when the 3D scale factor is given by $a(t)|_{t=0}=\sqrt{\tilde{c}_1}$.  This fact will become evident once we obtain the functional dependence of the 3D scale factor.

As we are interested in the behavior of the 3D scale factor for small $z$, (\ref{z4}) can be solved perturbatively by writing $z$ as a series solution of the form
\beq
z=z_0+z_1+...
\eeq
where $z_0$ is the solution to the linear approximation of (\ref{z4}), $z_1$ corresponds to the first-order correction term, etc.  Using this method and the definition of $z$ in (\ref{gamma}), we obtain the perturbative solution for the 3D scale factor of the form 
\bea\label{aspecial}
a(t)&=&\sqrt{\tilde{c}_1}\;\mbox{exp}\left[\tilde{\kappa}t^{(1-v)}-\frac{2(1-v)}{(2-v)}\tilde{\kappa}^2 t^{2(1-v)}+o(\tilde{\kappa}^3t^{3(1-v)})\right]\nonumber\\
&=&\sqrt{\tilde{c}_1}\left[1+\tilde{\kappa} t^{(1-v)}+\frac{(3v-2)}{2(2-v)}\tilde{\kappa}^2 t^{2(1-v)}+o(\tilde{\kappa}^3t^{3(1-v)})\right],
\eea
where we expanded the above expression for the 3D scale factor for small $\tilde{\kappa}t^{(1-v)}$, kept terms up to order $o(\tilde{\kappa}^2t^{2(1-v)})$, and used (\ref{ga}) with $\tilde{w}=1/3$.   Notice that $t$ is the time interval from when $a(t)|_{t=0}=\sqrt{\tilde{c}_1}$.

If one demands that the 3D scale factor is {\it real}, then we find the additional constraint that 
\beq
v\leq 0,
\eeq
where we used our definition of $\tilde{\kappa}$ in (\ref{tildekappa}) and the fact that $c_0>0$ and $\tilde{c}_1>0$ to obtain this result.  Consequently, when the higher-dimensional EoS parameter, $v$, is in fact negative,  the 3D scale factor exhibits \textit{accelerated expansion} for small time.  Remarkably,  a similar result emerges for the general case - that of accelerated expansion for small $t$ when $v<0$.  

Now inserting the solution for the 3D scale factor back into the 4D effective FRW field equations (\ref{rho2}) - (\ref{pd2}), we find a consistent solution up to and including order $o(\tilde{\kappa}^2t^{2(1-v)})$ when $\gamma_1$ is set equal to zero and 
\beq\label{constant1}
\frac{\rho_0}{\gamma_0^{(1+v)}}=\frac{3}{2}(-vc_0)^v.
\eeq	
It is interesting to note that (\ref{constant1}) is completely void of the integration constant $c_1$, where the value of $c_1$ sets the minimal size of the 3D scale factor.  Inserting (\ref{aspecial}) into (\ref{bfuna}) and setting $\gamma_1=0$, the higher-dimensional scale factor takes the form
\beq
b(t)=\frac{\gamma_0}{\sqrt[3]{\tilde{c}_1}}\cdot t\left[1-\frac{(4-v)}{(2-v)}\tilde{\kappa}t^{(1-v)}+o(\tilde{\kappa}^2t^{2(1-v)})\right]
\eeq
for small $\tilde{\kappa}t^{(1-v)}$ to order $o(\tilde{\kappa}t^{(1-v)})$.  Notice that the higher-dimensional scale factor is zero when $t=0$, which differs from that of the nonzero 3D scale factor for vanishing time.  Also notice that the higher-dimensional scale factor expands for small $t$ to lowest order.  

It should be noted that $\eta_1=0$ when the EoS parameters are given by $w=1/3,\;v=-1$.  Hence, for a universe with a 3D spatial component dominated by radiation, where the trace of the 4D energy momentum tensor is zero, one finds an early-time inflationary epoch.   Surprisingly, the results of the general case are remarkably similar to those of the special case.  In particular, we find that the general case also yields an early-time inflationary epoch, without the use of a scalar field.

\section{An exact solution: $v=0$}\label{v0}

In this appendix, we examine the behavior of the scale factors and the density for the special case of a vanishing higher-dimensional pressure.  Setting $v=0$, (\ref{pd2}) reduces to the form
\beq
0=\frac{\ddot a}{a}+\frac{\dot a^2}{a^2}\;,
\eeq
which yields an exact analytical solution for the 3D scale factor of the form
\beq
a(t)=a_0\;t^{1/2}.
\eeq
Inserting this solution back into the remaining FRW field equations, (\ref{rho2}) and (\ref{p2}), we find a consistent solution for the time dependence, with the coefficients subjected to the relation
\beq
a_0=\left(\frac{4\rho_0}{\eta_1\gamma_1}\right)^{1/4}.
\eeq
Notice that this expression for $a_0$ matches that of (\ref{a0}) when $v$ is set equal to zero.
If one demands that the 3D scale factor be \textit{real}, then we find that $\eta_1$ and $\gamma_1$ must be of the same sign.

Inserting our solution for the 3D scale factor into (\ref{rhofuna}), we find 
\beq\label{v0rho}
\rho(t)=\frac{\rho_0}{\gamma_1a_0^4}\cdot\frac{1}{t^2}\left[1-\left(\frac{6\gamma_0}{\eta_1\gamma_1}\right)a_0^{n-3}\;t^{-\eta_1/6}\right]^{-1}.
\eeq	
Now, plugging our solution for the 3D scale factor into (\ref{bfuna}), we find
\beq\label{v0b}
b(t)=\frac{\gamma_1}{a_0^n}\cdot\frac{1}{t^{n/2}}\left[1-\left(\frac{6\gamma_0}{\eta_1\gamma_1}\right)a_0^{n-3}\;t^{-\eta_1/6}\right].
\eeq
If one further demands that the density and the higher-dimensional scale factor remain positive for all time, we find the additional constraint that $\gamma_1>0$.  This requires $\eta_1>0$ for a real 3D scale factor.   Notice that for the exact solution of this appendix, the sign of $\gamma_0$ is left underdetermined.  If $\gamma_0$ is taken to be negative, then in the small time limit the 5D vacuum solutions of Chodos and Detweiler \cite{Chodos} are recovered.

Notice, however, that if $\gamma_0>0$, then there exists a time when the energy density diverges and the higher-dimensional scale factor approaches zero, explicitly 
\bea
\lim_{t\rightarrow t'}\;\rho(t)&\rightarrow&\infty\nonumber\\
\lim_{t\rightarrow t'}\;b(t)&\rightarrow&0\;\;\;\;\;\mbox{where}\;\;t'\equiv\left[\left(\frac{6\gamma_0}{\eta_1\gamma_1}\right)a_0^{n-3}\right]^{6/\eta_1}.
\eea
This behavior closely resembles that of the general case (see subsection \ref{z1}) near $z\sim1$, where we found a vanishing higher-dimensional scale factor and a divergent energy density when the shifted time approaches zero.

It should also be noted that for a universe with a 3D spatial component dominated by radiation, where the trace of the 4D energy momentum tensor is zero, the 3D EoS parameter is given by $w=1/3$, which yields $n=0$ when $v=0$.  For these values of the EoS parameters, we find that the higher-dimensional scale factor, given by (\ref{v0b}), asymptotes to a constant value for large time.

\section{An exact solution: $\tilde{w}=-1$}\label{w-1}

In this manuscript, the values of the EoS parameters $w$ and $v$ are constrained to lie within the range  $-1\leq w,v< 1$.  With this imposed constraint on our parameter space, the only values of the EoS parameters that yield an effective EoS parameter $\tilde{w}=-1$ are given by $w=v=-1$.   Notice that for these values, (\ref{rhofuna}) describes a constant energy density.

Using (\ref{effective}) to eliminate $\rho$ and $\tilde{p}$ from (\ref{rho2}) and (\ref{p2}), and rearranging, we obtain an expression of the form
\beq
\left[\gamma_1+\gamma_0\int a^{-4}\;dt\right]=\gamma_0a^{-4}\left[\frac{d}{dt}\left[\ln\left(\frac{\dot a}{a}\right)\right]\right]^{-1}.
\eeq
Now taking a derivative to eliminate the integral, rearranging, integrating twice, and inverting, the exact solution for the scale factor is found to be of the form
\beq
a(t)=a_0\cosh^{1/2}(\sqrt{2c}\;t)\;,
\eeq
where $c$ is a constant of integration and we set the arbitrary integration constant $t_0=0$.  Now, inserting our solution for the 3D scale factor into (\ref{bfuna}) and integrating, we obtain solutions for the scale factors of the form
\bea
a(t)&=&a_0\;\cosh^{1/2}\left(\sqrt{\frac{2\rho_0}{3}\;}\;t\right)\nonumber\\
b(t)&=&\frac{\gamma_0}{a_0^3}\sqrt{\frac{3}{2\rho_0}}\;\cosh^{1/2}\left(\sqrt{\frac{2\rho_0}{3}\;}\;t\right)\cdot\tanh\left(\sqrt{\frac{2\rho_0}{3}\;}\;t\right).
\eea
Notice that we evaluated the integration constant $c$ in terms of the constant energy density $\rho_0$ by inserting the general solution for the scale factors back into (\ref{rho0}) - (\ref{pd0}).   Also notice that when $t=0$, we find a vanishing higher-dimensional scale factor, yet a finite 3D scale factor.

\section{An exact solution: $w=v/3$}\label{wv/3}

For this unique relationship between the EoS parameters, (\ref{hyper}) reduces to a nonlinear algebraic equation of the form 
\beq
a^2\left(1-\frac{1}{c_2a^2}\right)=\left[-v\frac{c_0}{c_2}\right]^{v}t^{(1-v)},
\eeq
where we set the integration constant $t_0=0$. This equation can be solved exactly and yields a solution for the 3D scale factor of the form
\beq\label{asolvw/3}
a(t)=c_2^{-1/2}\left[1+2\kappa\;t^{(1-v)}\right]^{1/2}
\eeq
where we defined the quantity
\beq\label{kappa2}
\kappa\equiv\frac{1}{2}\left(-vc_0\right)^vc_2^{(1-v)},
\eeq
which agrees, by construction, with our definition of $\kappa$ in (\ref{kappa}) when $w=v/3$.  If we demand that the 3D scale factor be {\it real}, we find a constraint on the higher-dimensional EoS parameter of the form
\beq
v\leq 0,
\eeq
where we used our definition of $\kappa$ in (\ref{kappa2}) and the fact that $c_0>0$ to obtain this result.

Now inserting the solution for the 3D scale factor back into the 4D effective FRW field equations (\ref{rho2}) - (\ref{pd2}),  we find a consistent solution when $\gamma_1$ is set equal to zero and 
\beq\label{rho0wv/3}
\frac{\rho_0}{\gamma_0^{(1+v)}}=\frac{3}{2}(1-v)\cdot c_2(-vc_0)^v,
\eeq
which agrees precisely with (\ref{constant2}) when $w=v/3$.  Inserting (\ref{asolvw/3}) into (\ref{bfuna}) and setting $\gamma_1=0$, the higher-dimensional scale factor takes the form
\beq\label{bsolvw/3}
b(t)=\gamma_0 c_2^{3/2}\cdot t\left[1+2\kappa\;t^{(1-v)}\right]^{-1/2}.
\eeq
Notice that for $v=-1$, the higher-dimensional scale factor asymptotes to a constant for large time.  Inserting our solution for the 3D scale factor into (\ref{rhofuna}), we find an expression for the density of the form
\beq
\rho(t)=\frac{\rho_0}{\gamma_0^{(1+v)}}c_2^{-v}\cdot t^{-(1+v)}\left[1+2\kappa\;t^{(1-v)}\right]^{-1}.
\eeq
Notice that when $t$ is set equal to zero, the 3D scale factor yields a nonzero constant,  whereas the higher-dimensional scale factor vanishes and the density becomes infinite.  This behavior matches that of the perturbative solution of the general treatment for vanishing time found in subsection \ref{z1}.  Also notice that when (\ref{asolvw/3}) and (\ref{bsolvw/3}) are expanded for small $t$, they yield the perturbative solutions of (\ref{asol}) and (\ref{bsolgen}) for $v=w/3$, as one should expect.

We now wish to explore the scenario of accelerated expansion of the 3D scale factor for the special case of EoS parameters of this appendix.  Employing (\ref{asolvw/3}), we find an expression for the acceleration of the form
\beq\label{asolaccel}
\frac{\ddot a}{a}=-v(1-v)\kappa\;t^{-(1+v)}\left[1+2\kappa\;t^{(1-v)}\right]^{-2}\left[1+\frac{(1+v)}{v}\kappa\;t^{(1-v)}\right].
\eeq
Upon examination of (\ref{asolaccel}), we find accelerated expansion for the 3D scale factor when
\beq
v\kappa\cdot\left[1+\frac{(1+v)}{v}\kappa\;t^{(1-v)}\right]<0\;,
\eeq
when the higher-dimensional scale factor is constrained to lie within the range $-1\leq v< 1$.
Using the fact that $v<0$, to ensure that the 3D scale factor remains real, we find an epoch of accelerated expansion when
\beq\label{time}
t<\left[-3\frac{v(1-v)}{(1+v)}\cdot\frac{\gamma_0^{(1+v)}}{c_2^{-v}\rho_0}\right]^{1/(1-v)}\;\;\;\;\mbox{for}\;\;\;\;v\neq -1.
\eeq
where we used (\ref{rho0wv/3}) in obtaining (\ref{time}).  For the special case of $v=-1$, the last term in brackets in (\ref{asolaccel}) is unity and we find accelerated expansion for all time.  

To conclude this appendix, we found the exact analytical solution for the 3D and higher-dimensional scale factors for the case of $w=v/3$.  By demanding that the scale factors be real for all values of the EoS parameters, we find that the higher-dimensional pressure must in fact be negative.  This leads to an early-time epoch of accelerated expansion, which abruptly ends when (\ref{time}) is no longer satisfied.

\section{An exact solution: $w=0\;,\;v=-1$}\label{Aw0v-1}

In this appendix, we study the special case of a pressureless, matter-dominated 4D universe when the higher-dimensional EoS parameter is given by $v=-1$.  For this special case of the EoS parameters, (\ref{hyper}) takes the form of a nonlinear algebraic equation given by
\beq\label{cubic}
a^{3/2}\sqrt{1-\frac{1}{c_2a}\;}\cdot\left(1+\frac{2}{c_2a}\right)=\sqrt{\frac{3c_2}{c_0}}\;t
\eeq
where we set integration constant $t_0=0$.
\begin{figure}
\begin{center}$
\begin{array}{cc}
\includegraphics[scale=0.85]{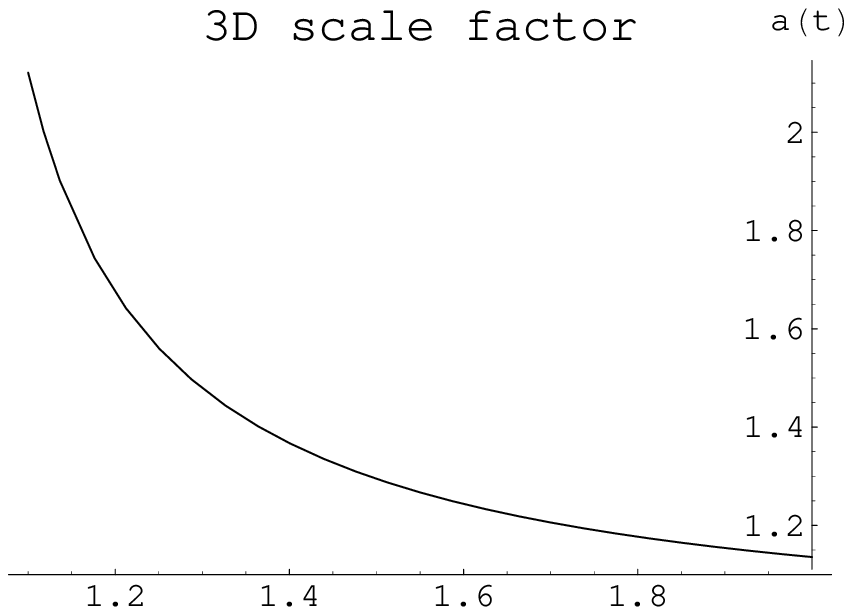}&
\includegraphics[scale=0.85]{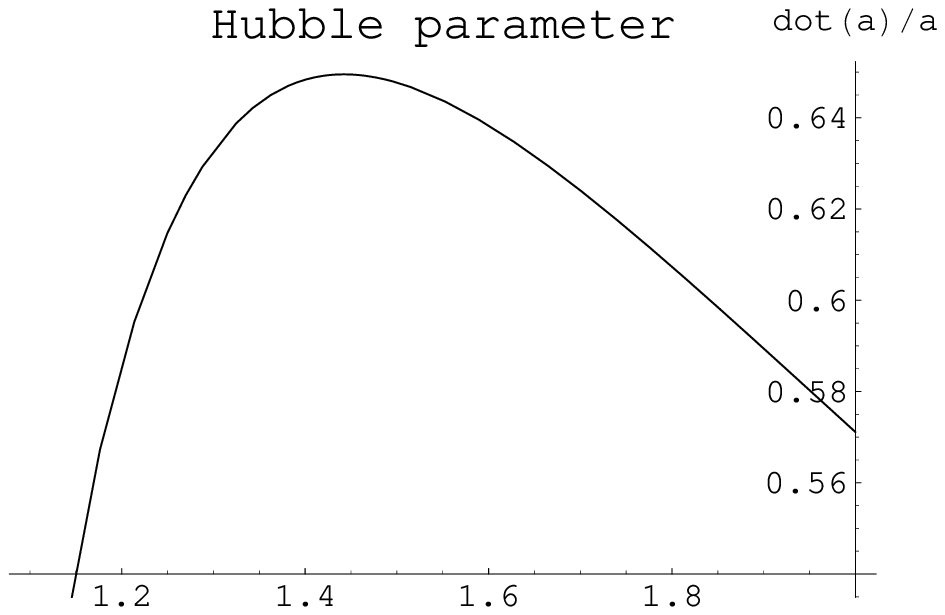}
\end{array}$
$
\begin{array}{c}
\includegraphics[scale=1.1]{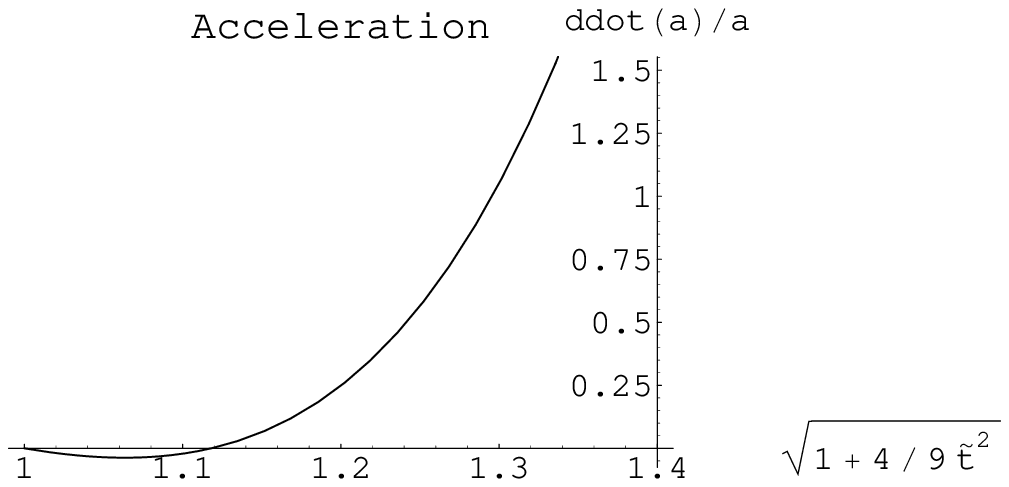}
\end{array}$
\caption{The plots are of the 3D scale factor, the Hubble parameter, and the acceleration vs. $\ell\equiv\sqrt{1+4/9\kappa t^2}$, respectively from  top to bottom, left to right.  Notice that all three of the embedded plots have $\ell$ on the horizontal axis, with the direction of decreasing $\ell$ corresponding to that of increasing $t$.  The plots were generated by arbitrarily setting the parameters $c_2=\kappa=1$.}  
\label{fig:3D}
\end{center}
\end{figure}

Notice that (\ref{cubic}) is {\it real} only when $a(t)>1/c_2$.   Now, squaring the above expression and performing some algebra, (\ref{cubic}) can be rewritten in the form of a cubic equation, which, in general, has three unique solutions \cite{table2}.  For the cubic equation of this appendix, we find one real root and a pair of complex conjugate roots. 
Disregarding the complex solutions, the real solution for the 3D scale factor is of the form
\beq\label{asolw0v-1}
a(t)=\frac{1}{c_2}\left[g+\frac{1}{g}-1\right],
\eeq
where we defined the quantity
\bea
g^3&\equiv& 1+\frac{9}{2}\kappa t^2\left[1+\sqrt{1+\frac{4}{9\kappa t^2}\;}\;\right]\;\;\;\;\mbox{where}\;\;\;\;\;\;\;\;\;\;\;\kappa\equiv\frac{c_2^4}{3c_0}>0\label{kappa3}\\
&=&\left(\frac{\ell+1}{\ell-1}\right)\hspace{3.9cm}\mbox{where}\;\;\;\;\;1<\ell\equiv\sqrt{1+\frac{4}{9\kappa t^2}\;}<\infty.\label{ell}
\eea
Notice that the real solution for the 3D scale factor, given by (\ref{asolw0v-1}), has a minimum value of $a(t)|_{t=0}=1/c_2$.  Also notice that the definition of $\kappa$ in (\ref{kappa3}) agrees with that of (\ref{kappa}) when $w=0,v=-1$ and is positive.  In (\ref{ell}), we have written the solution for the 3D scale factor in terms of the parameter $\ell$, which simplifies the analysis slightly.

Employing (\ref{asolw0v-1}), we obtain an expression for the Hubble parameter and the acceleration of the form
\bea\label{accelw0v-1}
\frac{\dot a}{a}&=&\sqrt{\kappa}\;\frac{\sqrt{\ell^2-1}}{\ell}\left[\left(\frac{\ell+1}{\ell-1}\right)^{1/3}-\left(\frac{\ell-1}{\ell+1}\right)^{1/3}\right]\left[\left(\frac{\ell+1}{\ell-1}\right)^{1/3}+\left(\frac{\ell-1}{\ell+1}\right)^{1/3}-1\right]^{-1}\nonumber\\
\frac{\ddot a}{a}&=&\frac{\kappa}{2}\;\frac{(\ell^2-1)}{\ell^3}\left[(2\ell-3)\left(\frac{\ell+1}{\ell-1}\right)^{1/3}+(2\ell+3)\left(\frac{\ell-1}{\ell+1}\right)^{1/3}\right]\cdot\left[\left(\frac{\ell+1}{\ell-1}\right)^{1/3}+\left(\frac{\ell-1}{\ell+1}\right)^{1/3}-1\right]^{-1}.\nonumber\\
\eea
Figure \ref{fig:3D} shows plots of the 3D scale factor, the Hubble parameter, and the acceleration vs the parameter $\ell$, where we arbitrarily set  the parameters $c_2=\kappa=1$ in generating the plots.  The direction of decreasing $\ell$ corresponds to that of increasing time, with $\ell\rightarrow 1$ as $t\rightarrow\infty$. 

Inserting (\ref{asolw0v-1}) and (\ref{accelw0v-1}) back into the 4D effective FRW field equations, (\ref{rho2}) - (\ref{pd2}), we find a consistent solution when
\beq\label{rho0w0v-1}
\rho_0=\frac{2c_2}{c_0},
\eeq
which agrees precisely with (\ref{constant2}) for $w=0,v=-1$. 

Upon examination of (\ref{accelw0v-1}), we find an early-time (large $\ell$) epoch of accelerated expansion for the 3D scale factor.  This epoch of accelerated expansion occurs when
\beq\label{time2}
t<\frac{4}{c_2^{3/2}}\sqrt{\frac{2}{3\rho_0}}\;,
\eeq
where we used (\ref{kappa3}) - (\ref{rho0w0v-1}) to obtain this result.

To conclude this appendx, we found the real, exact analytical solution for the 3D scale factor for the special case of EoS parameters given by $w=0, v=-1$.  This case corresponds to a pressureless, matter-dominated 4D universe with the higher-dimensional EoS parameter set to $v=-1$.  The solution for the 3D scale factor leads to an early-time epoch of accelerated expansion, which abruptly ends when (\ref{time2}) is no longer satisfied.


\begin{thebibliography}{999}
\bibitem{KK}
Th. Kaluza, Sitzungseber. Press. Akad. Wiss. Phys. Math. Klasse 996 (1921).

O. Klein, Z.F. Physik {\bf 37}, 895 (1926).

O. Klein, Nature {\bf 118}, 516 (1926).

\bibitem{Apple}
T. Applequist, A. Chodos, P.G.O. Frend, {\it Modern Kaluza-Klein theories}, (Addison Wesley, 1987).

P.S. Wesson, {\it Five-dimensional physics: classical and quantum consequences of Kaluza-Klein cosmology}, (World Scientific, 2006).

\bibitem{Large}
N. Arkani-Hamed, S. Dimopoulos, G. Dvali, 
Phys. Lett. B \textbf{429}, 263 (1998), (hep-ph/9803315v1).

I. Antoniadis, N. Arkani-Hamed, S. Dimopoulos, G. Dvali, 
Phys. Lett. B \textbf{436}, 257 (1998), (hep-ph/9804398v1).

N. Arkani-Hamed, S. Dimopoulous, G. Dvali, 
Phys. Rev. D \textbf{59}, 086004 (1999), (hep-ph/9807344v1).


L. Randall, R. Sundrum, 
Phys. Rev. Lett. \textbf{83}, 4690 (1999), (hep-th/9906064).

G. Dvali, G. Gabadadze, M. Porrati, 
Phys. Lett. B  \textbf{485}, 208 (2000), (hep-th/0005016).

G. Dvali, G. Gabdadadze, 
Phys. Rev. D \textbf{63} 065007 (2001), (hep-th/0008054).


\bibitem{Accel}
S. Perlmutter et al., 
Astrophys. J. \textbf{517}, 565 (1999).

A. Riess et al., 
Astron. J. \textbf{116}, 1009 (1998), (astro-ph/9805201).

\bibitem{CMB}
C. L. Bennett et al., 
Astrophys. J. Suppl \textbf{148}, 1 (2003), (astro-ph/0302207v3).

C. B. Netterfield et al., 
Astrophys. J. \textbf{571}, 604 (2002), (astro-ph/0104460v3).

N. W. Halverson et al., 
Astrophys. J. \textbf{568}, 38 (2002), (astro-ph/0104489v2).

\bibitem{CosCon}
S. Weinberg, 
Rev. Mod. Phys. \textbf{61}, 1 (1989)


\bibitem{Inflation}
Alan H. Guth, 
Phys. Rev. D \textbf{23}, 347 (1981)
  
 \bibitem{Liddle}
Andrew R. Liddle and David H. Lyth,  \textit{Cosmological Inflation and Large-Scale Structure}, (Academic Press, Inc., 2000).
 



\bibitem{Mohammedi}
N. Mohammedi, 
Phys.Rev. D \textbf{65}, 104018 (2002)  (hep-th/0202119).




\bibitem{others}
E. A. Leon, J. A. Nieto, R. Nunez-Lopez, A. Lipovka, 
Mod. Phys. Lett. A \textbf{26}, 805 (2011), (arXiv:1012.3556 [gr-qc]).

D. Panigrahi, S. Chatterjee, 
Grav. Cosmol. \textbf{17}, 18 (2011), (arXiv:1006.0476v2 [gr-qc]).

F. Darabi, \textit{Trends in General Relativity and Quantum Cosmology}, (Nova Science Publishers, New York 2006), (arXiv:0902.1863v1 [gr-qc]).

F. Darabi, 
Mod. Phys. Lett. A \textbf{25}, No. 19,  1635 (2010), (arXiv:0901.0835v3 [gr-qc]).
 
Masato Ito, 
(arXiv:0903.5118v2 [hep-th]).

Andrew, K., Bolen, B., Middleton, C. A., 
Grav. and Gen. Rel., \textbf{39}, 2061 (2007), (arXiv:0708.0373 [gr-qc]).

Torsten Bringmann, Martin Eriksson, Michael Gustafsson, 
Phys. Rev. D  \textbf{68}, 063516 (2003), (arXiv:astro-ph/0303497v2).

F. Darabi, 
Class. Quant. Grav. \textbf{20}, 3385 (2003), (arXiv: gr-qc/0301075v2).

\bibitem{Pan}
D. Panigrahi, Y. Z. Zhang,  S.  Chatterjee, 
Int. J. Mod. Phys. A \textbf{21}, 6491 (2006), (arXiv:gr-qc/0604079v2).

\bibitem{Chodos}
A. Chodos, S. Detweiler, 
Phys. Rev. D \textbf{21}, 2167 (1980)


\bibitem{table}
I.S. Gradshteyn and I.M. Ryzhik, \textit{Table of Integrals, Series, and Products}, (Academic Press, Inc., 1965).

\bibitem{table2}
Milton Abramowitz and Irene A. Stegun, \textit{Handbook of Mathematical Functions with Formulas, Graphs, and Mathematical Tables}, (Dover Publications, Inc., 1965).


\end{thebibliography}
 \end{document}